\documentclass[prb,aps,twocolumn,amsmath,amssymb,floatfix,superscriptaddress]{revtex4}
\usepackage[dvips]{graphics}
\usepackage{color}
\definecolor{dred}{rgb}{0,0,0.6}
\usepackage{amsmath}

\begin{document}

\title{Circular current in a one-dimensional open quantum ring in the presence
of magnetic field and spin-orbit interaction}

\author{Moumita Patra}

\affiliation{School of Physical Sciences, Indian Association for the Cultivation of Science, 2A \& 2B Raja S. C.
Mullick Road, Kolkata 700032, India}

\begin{abstract}

In an open quantum system having a channel in the form of loop geometry, the current inside the channel, namely
circular current, and overall junction current, namely transport current, can be different. A quantum ring has
doubly degenerate eigen energies due to periodic boundary condition that is broken in an asymmetric ring where
the ring is asymmetrically connected to the external electrodes. Kramers' degeneracy and spin degeneracy can
be lifted by considering non-zero magnetic field and spin-orbit interaction (SOI), respectively. Here, we find that
symmetry breaking impacts the circular current density vs energy ($E$) spectra in addition to lifting the degeneracy.
For charge and spin current densities, the corresponding effects are not the same. Under symmetry-breaking they
may remain symmetric or anti-symmetric or asymmetric around $E = 0$ whereas the transmission function (which is
proportional to the junction current density) vs energy characteristic remains symmetric around $E = 0$.
This study leads us to estimate the qualitative nature of the circular current and the choices of
Fermi-energy/chemical potential to have a net non-zero current. As a result, we may manipulate the system to
generate pure currents of charge, spin, or both, which is necessary for any spintronic and electronic applications.

\end{abstract}

\maketitle

\section{Introduction}

\begin{figure}[ht]
{\centering \resizebox*{8cm}{4cm}{\includegraphics{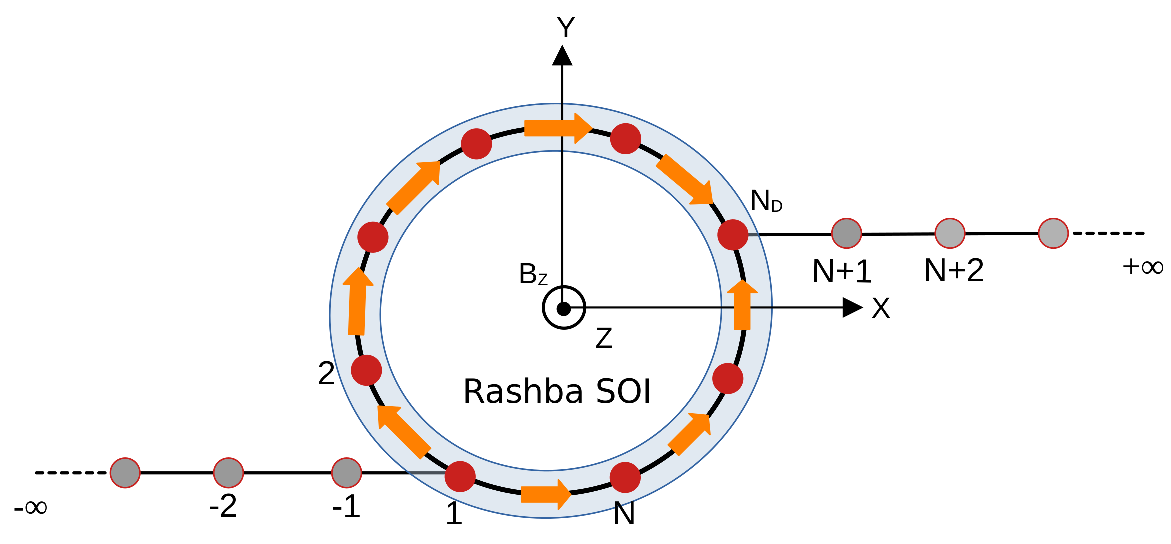}}\par}
\caption{(Color online). Schematic representation of a ring scatterer in the presence
of external magnetic field $B_Z$ along $Z\,$direction and Rashba SOI in the $XY$ plane, where the
ring is confined. The magnitudes and directions of each bond current are
indicated by the arrows. The ring is attached to two semi-infinite metallic electrodes, namely
source and drain. The ring contains $N$ atomic-sites. The source is connected to the first site
of the ring and we have attached the drain at the $N_D\,$-th site.}
\label{f2}
\end{figure}

\begin{figure*}
{\centering \resizebox*{11.5cm}{2.5cm}{\includegraphics{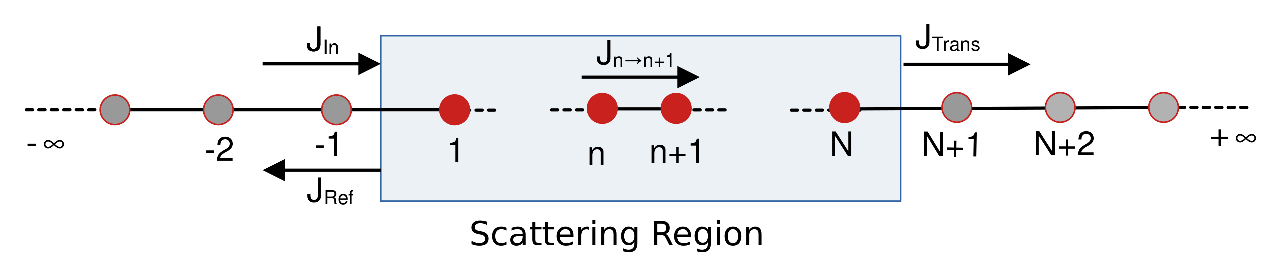}}\par}
\caption{(Color online). The tight-binding model to describe the scattering principle, where we consider
a finite-size scattering region sandwiched between an infinite, one-dimensional linear conductor.
$J_{\mbox{\tiny{In}}}(E)$, $J_{\mbox{\tiny{Ref}}}(E)$, and $J_{\mbox{\tiny{In}}}(E)$ are the incident,
reflected and transmitted current (or the overall junction current) densities, respectively. Within the
scattering region, $J_{n\rightarrow n+1}(E)$ is the bond current between any successive sites $n$ and $n+1$.}
\label{f1}
\end{figure*}

The current inside the channel and the overall junction current have substantially distinct quantitative and
qualitative characteristics~\cite{cir1a,cir2,cir3,cir4,Nitzan1,cir6,SKM,cir8,cir9,cir10} for a quantum junction
with a channel shaped like a loop or ring~\cite{ring1,ring2,ring3,ring4}. Let us consider
the open quantum junction, having a ring channel as shown in Fig.~\ref{f2}. Here, the quantum ring is
connected to the incoming (namely, source) and outgoing (namely, drain) electrodes.
The zero-temperature bond current within the ring and overall junction current can be written by the
Landauer formula as:
\begin{equation}
I_{n \rightarrow n+1}(V) = \int\limits_{E_F - \frac{eV}{2}}^{E_F+\frac{eV}{2}}J_{n \rightarrow n+1}(E) \, dE
\label{j4}
\end{equation}
\noindent
and
\begin{equation}
I_{T}(V) = \frac{2e}{h} \int\limits_{E_F-\frac{eV}{2}}^{E_F+\frac{eV}{2}}T(E) \, dE,
\label{eq2a}
\end{equation}
\noindent
respectively~\cite{mpprb}. Here $e$ is the electronic charge and $h$ is the
Planck’s constant. $J_{n \rightarrow n + 1}(E)$ and $T(E)$ are the circular current density
and the transmission function, respectively. $E_F$ is the equilibrium Fermi-function or the chemical potential.
For a linear conductor (as shown in Fig.~\ref{f1})
as these two currents should be equal due to the conservation of charge. Therefore,
\begin{equation}
J_{i \rightarrow i+1} (E) = 2T(E)
\label{linear}
\end{equation}
\noindent for a linear conductor. In this paper, we choose $e = h =1$.

In a quantum junction with ring geometry (as shown in Fig.~\ref{f2})
the currents in the upper and lower arms of the ring are not the same~\cite{jay1,jay2}. But
as the charge current is a conserved quantity, thus each bond in an arm carries the same current.
Let $I_C^U$ and $I_C^L$ be the current in the upper and the lower arms of the ring, respectively
such that each bond in the upper or lower arm carries $I_C^U$ or $I_C^L$ current, respectively.
Due to the current conservation, for a ring junction, we can write
\begin{equation}
I_{T_C} = I_{C}^U - I_C^L.
\label{kir}
\end{equation}
\noindent
$I_{T_C}$ is the junction charge current. Here we assign a positive sign to the current flowing in the
clock-wise direction. From Eq.~\ref{kir}, we can see that the currents inside and outside a ring-channel are not
the same. The junction transmission
function $T_{C}(E)$ is always a positive quantity with upper bound 2 ($=1+1$ due to contribution of the
up and down spins). Whereas $J_{i \rightarrow i+1} (E)$ may be positive or negative, with no bound for a ring channel.
The circular current is defined as the average of all the bonds current, i.e.,
\begin{equation}
I = \frac{\displaystyle\sum_n l_{n\rightarrow n+1}
I_{n\rightarrow n+1}}{\displaystyle\sum_n l_{n\rightarrow n+1}};
\label{cireq}
\end{equation}
\noindent
$l_{n\rightarrow n+1}$ is the bond-length. $\displaystyle\sum_n l_{n\rightarrow n+1}$ is the length of the ring $L$.
For a ring with fixed bond-length $a$ and $N$ number of atomic sites, $L = Na$. Thus we can rewrite Eq.~\ref{cireq} as
\begin{equation}
I = \frac{1}{N}\sum_n I_{n\rightarrow n+1}.
\label{cireq1}
\end{equation}
We can determine the characteristics of the circular current inside the channel
(using Eq.~\ref{j4}) and the overall junction current (using Eq.~\ref{eq2a})
by studying circular current density and transmission function as a function of energy, respectively.

A quantum ring has doubly degenerate energy states due to its periodic boundary condition $N \equiv N+1$,
which implies that the energy eigenstates, with momenta $+k$ and $-k$ have same energy eigen values.
This degeneracy can be broken if we connect the electrodes to the ring in such a way, that the length of the upper
arm and lower arm are not the same. We call it asymmetric ring. According to the Kramers' degeneracy
theorem~\cite{kra1,kra2,kra3}, a Fermionic system has at least two fold degeneracy if it remains unchanged under
time-reversal (TR) transformation which is defined as
$$\mbox{\boldmath$T$} :t \mapsto -t;$$
\noindent such that $[\mbox{\boldmath$T$},~\mbox{\boldmath$H$}] = 0$.
TR symmetry can be broken in the presence of magnetic flux $\phi$. In this situation
$E(k + \phi, \uparrow) \neq E(-k + \phi, \uparrow)$.  The spin-degeneracy can be lifted by adding spin-orbit
interaction~\cite{spind}. In the presence of SOI, we can write $E(k, \uparrow) \neq E(k, \downarrow)$.
Whereas in the presence of both $\phi$ and SOI, we can write $E(k + \phi, \uparrow) \neq E(-k + \phi, \downarrow)$.

In this paper, we investigate the circular current density and transmission function as a function
of energy in the presence of an external
magnetic-field, spin-orbit interaction, and both. Here we find that while the transmission function
spectra remain symmetric around around $E = 0$, the symmetry-breaking
has impact on circular current density vs energy characteristics. For a symmetric-ring (where the
ring is symmetrically connected to the electrodes), we show that the charge current density $J_C(E)$
is zero without any external fields and in the presence of SOI. $J_C(E)$
is non-zero in the presence of magnetic-field and both SOI and magnetic flux $\phi$.
The circular spin-current density is zero in the absence of SOI and non-zero in the other cases.
The non-zero $J_C (E)$ and $J_S (E)$ are anti-symmetric around $E = 0$. Thus following Eq.~\ref{j4},
we can write in a symmetric-ring,
the net circular charge and spin currents become 0 if we set the Fermi-energy $E_F$ at 0.
In asymmetric-ring, $J_C (E)$ is always non-zero. It is symmetric around $E = 0$ in the absence
of $\phi$, and asymmetric for non-zero magnetic-field. Thus net circular charge current $I_C$ is
always finite in a asymmetric-ring for any $E_F$. In an asymmetric-ring, $J_S (E)$ is non-zero
in the presence of SOI (similar to symmetric ring). In the presence of SOI only, $J_S (E)$ is
anti-symmetric around $E = 0$, thus net $I_S$ becomes zero for $E_F = 0$. Whereas $J_S (E)$ is
asymmetric around $E = 0$, in the presence of both $\phi$ and SOI, thus this becomes finite for
any $E_F$. The overall junction charge transmission function
$T_C (E)$ is non-zero in the presence and absence of the interactions. $T_S (E)$ is non-zero in
the presence of both $\phi$ and SOI only.
As they are always symmetric around $E = 0$, the net $I_{T_C}$ (and $I_{T_S}$ in the presence
of both $\phi$ and SOI) are finite for any choice of $E_F$.

These findings lead us to manipulate the system to generate pure charge current or pure spin current
or both which is very important for designing electronic and spintronic
devices~\cite{dev1,th1,th2,th3,th4,th5,th6,th7,th8,th9,th10}. For example, in the presence of SOI,
when unpolarized electrons are injected in a symmetric-ring, for any non-zero $E_F$ a pure spin current
(where $I_C$ is zero) is generated within the ring. On the other hand, in an asymmetric-ring, for $E_F = 0$,
pure charge current (where $I_S$ is zero) is produced and for $E_F \neq 0$ both $I_C$ and $I_S$ become finite. 

The outline of the paper is as follows. First, in section II, we define the general model and describe the
scattering formalism to estimate the current inside the scatterer and overall junction current.
In section III we calculate all the components of circular current densities and transmission functions
in the absence and presence of magnetic field, SOI, and both considering symmetric
and asymmetric ring. Finally we conclude with a discussion in section IV.

\section{Theory}

We follow the scattering formalism~\cite{scatter1}, where we can define the circular currents densities and transmission
function of an open quantum system by wave amplitudes. Let us consider Fig.~\ref{f1} where
a Bloch wave incidents with energy $E$ from the left and transmitted to the right. In between, we consider
a tight-binding~\cite{tb} scatterer (shown by blue color). The Hamiltonian for the entire system
is considered as:
\begin{equation}
\mbox{\boldmath $H$} = \mbox{\boldmath $H$}_S + \mbox{\boldmath $H$}_{\mbox{\tiny Scatterer}} +
\mbox{\boldmath $H$}_D + \mbox{\boldmath $H$}_C.
\label{eq1}
\end{equation}
\noindent Here,
\begin{equation}
\mbox{\boldmath $H$}_{S/D} = \displaystyle\sum\limits_{n = -1/N+1}^{-\infty/+\infty}
\left(\mbox{\boldmath $c$}_{n+1}^\dagger\mbox{\boldmath $t$}_{0}^\dagger\mbox{\boldmath $c$}_n
+ \mbox{\boldmath $c$}_{n}^\dagger\mbox{\boldmath $t$}_{0}\mbox{\boldmath $c$}_{n+1} \right),
\label{eq2}
\end{equation}
\begin{equation}
\mbox{\boldmath $H$}_{\mbox{\tiny Scatterer}} = \displaystyle\sum\limits_{n = 1}^{N}
\left(\mbox{\boldmath $c$}_{n+1}^\dagger\mbox{\boldmath $t$}_{n\rightarrow n+1}^\dagger\mbox{\boldmath $c$}_n
+ \mbox{\boldmath $c$}_{n}^\dagger\mbox{\boldmath $t$}_{n\rightarrow n+1}\mbox{\boldmath $c$}_{n+1} \right),
\label{eq3}
\end{equation}
\begin{eqnarray}
\mbox{\boldmath $H$}_{C} & = &
\left(\mbox{\boldmath $c$}_{-1}^\dagger\mbox{\boldmath $t$}_{-1\rightarrow 1}^\dagger\mbox{\boldmath $c$}_1
+ \mbox{\boldmath $c$}_{1}^\dagger\mbox{\boldmath $t$}_{-1\rightarrow 1}\mbox{\boldmath $c$}_{-1} \right) \nonumber \\
& + & \left(\mbox{\boldmath $c$}_{N}^\dagger\mbox{\boldmath $t$}_{N\rightarrow N+1}^\dagger\mbox{\boldmath $c$}_{N+1}
+ \mbox{\boldmath $c$}_{N+1}^\dagger\mbox{\boldmath $t$}_{N\rightarrow N+1}\mbox{\boldmath $c$}_{N} \right). \nonumber \\
\label{eq3a}
\end{eqnarray}
\noindent
$\mbox{\boldmath $H$}_S$, $\mbox{\boldmath $H$}_{\mbox{\tiny Scatterer}}$, $\mbox{\boldmath $H$}_D$ and
$\mbox{\boldmath $H$}_C$ are the sub-Hamiltonian representing the left side of scatterer i.e., the source, the scatterer,
right side of scatterer i.e., the drain, and their coupling, respectively. Here
\noindent
$\mbox{\boldmath $c$}_n^\dagger=\left(\begin{array}{cc}
    c_{\uparrow,n}^\dagger & c_{\downarrow,n}^\dagger
\end{array}\right)$. $c_{\sigma,n}^\dagger$ ($\sigma = \uparrow, \downarrow$) is the creation operator.
$\mbox{\boldmath $c$}_n=\left(\begin{array}{cc}
    c_{\uparrow,n} \\ c_{\downarrow,n}
\end{array}\right)$. $c_{\sigma,n}$ ($\sigma = \uparrow, \downarrow$) is the annihilation operator.
$\mbox{\boldmath $t$}_{\sigma\sigma', n\rightarrow n+1}=\left(\begin{array}{cc}
    t_{\uparrow\uparrow, n \rightarrow n+1} & t_{\uparrow\downarrow, n \rightarrow n+1}\\
    t_{\downarrow\uparrow, n \rightarrow n+1} & t_{\downarrow\downarrow, n\rightarrow n+1}
\end{array}\right)$. $t_{\sigma\sigma', n \rightarrow n+1}$ is the nearest neighbor coupling for an electron with spin
$\sigma$ and hopped as spin $\sigma'$ ($\sigma, \sigma' = \uparrow, \downarrow$). $\mbox{\boldmath $t$}_0$
is the hopping matrix for the left and right side of the scatterer where
$\mbox{\boldmath $t$}_0=\left(\begin{array}{cc}
    t_0 & 0\\
    0 & t_0
\end{array}\right)$.

When up spin incidents with energy $E = 2 t_0\cos(ka)$, the solution for the left of the scattering region can be
written in terms of incoming and reflected waves as
\begin{eqnarray}
|\mbox{\boldmath $\Psi$}_{\uparrow,L}\rangle & = & \sum_{n \leq -1}\left(\begin{array}{cc}
A_{\uparrow} e^{i(n + 1)ka} + B_{\uparrow\uparrow}e^{-i(n + 1)ka} \\
B_{\uparrow\downarrow}e^{-i(n + 1)ka}\end{array}\right)|n, \uparrow\rangle \nonumber \\
&&\label{eq4a}
\end{eqnarray}
\noindent
Whereas if down spin incidents from left, the wave function looks that
\begin{eqnarray}
|\mbox{\boldmath $\Psi$}_{\downarrow,L}\rangle & = & \displaystyle\sum_{n \leq -1}\left(\begin{array}{cc}
B_{\downarrow\uparrow}e^{-i(n + 1)ka} \\
A_{\downarrow} e^{i(n + 1)ka} + B_{\downarrow\downarrow}e^{-i(n + 1)ka}\end{array}\right)|n, \downarrow\rangle \nonumber \\
\label{eq4b}
\end{eqnarray}
\noindent
$A_{\sigma}$ = Incident amplitude of an electron with spin $\sigma$ ($\sigma = \uparrow, \downarrow$).\\
\noindent
$B_{\sigma\sigma'}$ = Reflection amplitude of a electron which is incident with spin $\sigma$ and reflects as $\sigma'$
($\sigma,\sigma' = \uparrow, \downarrow$).

The solution to the right is given by a transmitted wave in the case of up spin incidence as
\begin{eqnarray}
|\mbox{\boldmath $\Psi$}_{\uparrow,R}\rangle & = & \displaystyle\sum_{n > N}\left(\begin{array}{cc}
\tau_{\uparrow\uparrow}e^{inka} \\
\tau_{\uparrow\downarrow} e^{inka} \end{array}\right)|n, \downarrow\rangle
\label{eq5a}
\end{eqnarray}
\noindent
Similarly for the down spin incidence, the solution is
\begin{eqnarray}
|\mbox{\boldmath $\Psi$}_{\downarrow,R}\rangle & = & \displaystyle\sum_{n > N}\left(\begin{array}{cc}
\tau_{\uparrow\downarrow}e^{inka} \\
\tau_{\downarrow\downarrow} e^{inka} \end{array}\right)|n, \downarrow\rangle.
\label{eq5b}
\end{eqnarray}
\noindent
$\tau_{\sigma\sigma'}$ = Transmission amplitude of an electron with spin $\sigma$ and
transmitted as spin $\sigma'$ ($\sigma, \sigma' = \uparrow, \downarrow$).

For the scattering region the steady-state wave function for up and down spins have the form,
\begin{equation}
|\mbox{\boldmath $\Psi$}_{\uparrow,S}\rangle  = \displaystyle\sum_{n = 1}^{N}\left(\begin{array}{cc}
C_{\uparrow\uparrow,n} \\
C_{\uparrow\downarrow,n}\end{array}\right)|n, \downarrow\rangle,
\label{eq6a}
\end{equation}
\noindent
\begin{eqnarray}
|\mbox{\boldmath $\Psi$}_{\downarrow,S}\rangle = \displaystyle\sum_{n = 1}^{N}\left(\begin{array}{cc}
C_{\downarrow\uparrow,n}\\
C_{\downarrow\downarrow,n}\end{array}\right)|n, \downarrow\rangle,
\label{eq6b}
\end{eqnarray}
\noindent respectively. $C_{\sigma\sigma',n}$ ($\sigma,~\sigma' = \uparrow,\downarrow$) are the
wave-amplitudes. The current density between any two neighboring-sites of the scatterer, that is
$n$ and $n+1$ can be evaluated as
\begin{eqnarray}
J_{\sigma\sigma',n\rightarrow n+1} (E) & = & \frac{e}{\hbar}\Big(C_{\sigma\sigma',n}^\dagger
t_{\sigma\sigma',n\rightarrow n+1}^*C_{n+1} \nonumber \\
& - & C_{n+1}^* t_{\sigma\sigma',n \rightarrow n+1}C_n\Big).
\label{eq7}
\end{eqnarray}
$\hbar$ is the reduced Planck's constant; Net up and down spin current densities with in a bond are defined as
\begin{eqnarray}
J_{\uparrow, n\rightarrow n+1}(E) & = & J_{\uparrow\uparrow, n\rightarrow n+1}(E) +
J_{\downarrow\uparrow, n\rightarrow n+1}(E)\nonumber \\
J_{\downarrow, n\rightarrow n+1}(E) & = & J_{\uparrow\downarrow, n\rightarrow n+1}(E) +
J_{\downarrow\downarrow, n\rightarrow n+1}(E),
\label{eq7a}
\end{eqnarray}
\noindent respectively. The net charge and spin current densities are given by
\begin{eqnarray}
J_{C, n\rightarrow n+1}(E) & = & J_{\uparrow, n\rightarrow n+1}(E) +
J_{\downarrow, n\rightarrow n+1}(E)\nonumber \\
J_{S, n\rightarrow n+1}(E) & = & J_{\uparrow, n\rightarrow n+1}(E) -
J_{\downarrow, n\rightarrow n+1}(E),
\label{eqba}
\end{eqnarray}
\noindent respectively. The net current density on the left side of the scatter is
\begin{equation}
J_{\sigma\sigma', -2\rightarrow-1}(E) =  J_{\mbox{\tiny In}_{\sigma\sigma'}}(E) - J_{\mbox{\tiny Ref}_{\sigma\sigma'}}(E)
\label{eq7c}
\end{equation}
\noindent
$J_{\mbox{\tiny In}_{\sigma\sigma'}}(E)$, $J_{\mbox{\tiny Ref}_{\sigma\sigma'}}(E)$ are the incident and reflected currents
density respectively. The spin dependent incident, reflected, and transmitted current density are given by
\begin{eqnarray}
J_{\mbox{\tiny In}_{\sigma}}(E) & = & \frac{\Gamma}{\hbar}|A_\sigma|^2, \nonumber \\
J_{\mbox{\tiny Ref}_{\sigma\sigma'}}(E) & = & \frac{\Gamma}{\hbar}|B_{\sigma\sigma'}|^2, \nonumber \\
J_{\mbox{\tiny Trans}_{\sigma\sigma'}}(E) & = & \frac{\Gamma}{\hbar}|\tau_{\sigma\sigma'}|^2,
\label{eq7d}
\end{eqnarray}
\noindent respectively. $\Gamma(E) = 2 |t_0 \sin(ka)|$. $\sigma,~\sigma' = \uparrow,\downarrow$. 
The reflection and transmission functions are defined as
\begin{equation}
R_{\sigma\sigma'}(E) = |B_{\sigma\sigma'}|^2
\label{reff}
\end{equation}
\begin{equation} 
T_{\sigma\sigma'}(E) = |\tau_{\sigma\sigma'}|^2,
\label{trans}
\end{equation}
\noindent respectively. $\sigma, \sigma' = \uparrow, \downarrow$. Therefore,
\begin{eqnarray}
R_{\uparrow}(E) & = & R_{\uparrow\uparrow}(E) + R_{\downarrow\uparrow}(E); \nonumber \\
R_{\downarrow}(E) & = & R_{\uparrow\downarrow}(E) + R_{\downarrow\downarrow}(E); \nonumber \\
T_{\uparrow}(E) & = & T_{\uparrow\uparrow}(E) + T_{\downarrow\uparrow}(E); \nonumber \\
T_{\downarrow}(E) & = & R_{\uparrow\downarrow}(E) + T_{\downarrow\downarrow}(E).
\label{rt}
\end{eqnarray}
Once the reflection and transmission functions are known, one can show the conservation condition
$(R_{\uparrow}(E) + R_{\downarrow}(E)) + (T_{\uparrow}(E) + T_{\downarrow}(E)) = 2$. The charge and
spin transmission functions are defined as
\begin{eqnarray}
T_C(E) & = & T_{\uparrow}(E) + T_{\downarrow}(E); \nonumber \\
T_S(E) & = & T_{\uparrow}(E) - T_{\downarrow}(E),
\label{tcs}
\end{eqnarray}
\noindent respectively.

\section{Results and discussion}

Now we discuss the results of considering a ring scatterer. All the results are computed for ring size $N = 6$ where
the source is always connected to the first site of the ring, i.e.,
$N_S = 1$. This is for the better representation of the results though this qualitative analysis is true for any $N$.
Throughout the calculations we choose the hopping parameters for the electrodes are $2\,$eV and all other hoppings are $1\,$eV.
The inter-atomic spacing ($a$) is considered to be $10\,$nm. 
Now we discus how the current density as well as the transmission function vs energy
spectra are affected by the magnetic field, SOI, and both.

\subsection{Transport through a perfect ring without any external field}

Let us first understand the open-quantum transport considering a perfect ring scatterer which has
identical nearest-neighbor hopping integral. The carriers have two possible
paths namely, upper or lower arms of the ring to travel from source to drain.
Let us consider Fig.~\ref{f2}. The path between site-1 (where the source is connected to the ring) to $N_D$
\begin{figure}[ht]
{\centering \resizebox*{8cm}{6cm}{\includegraphics{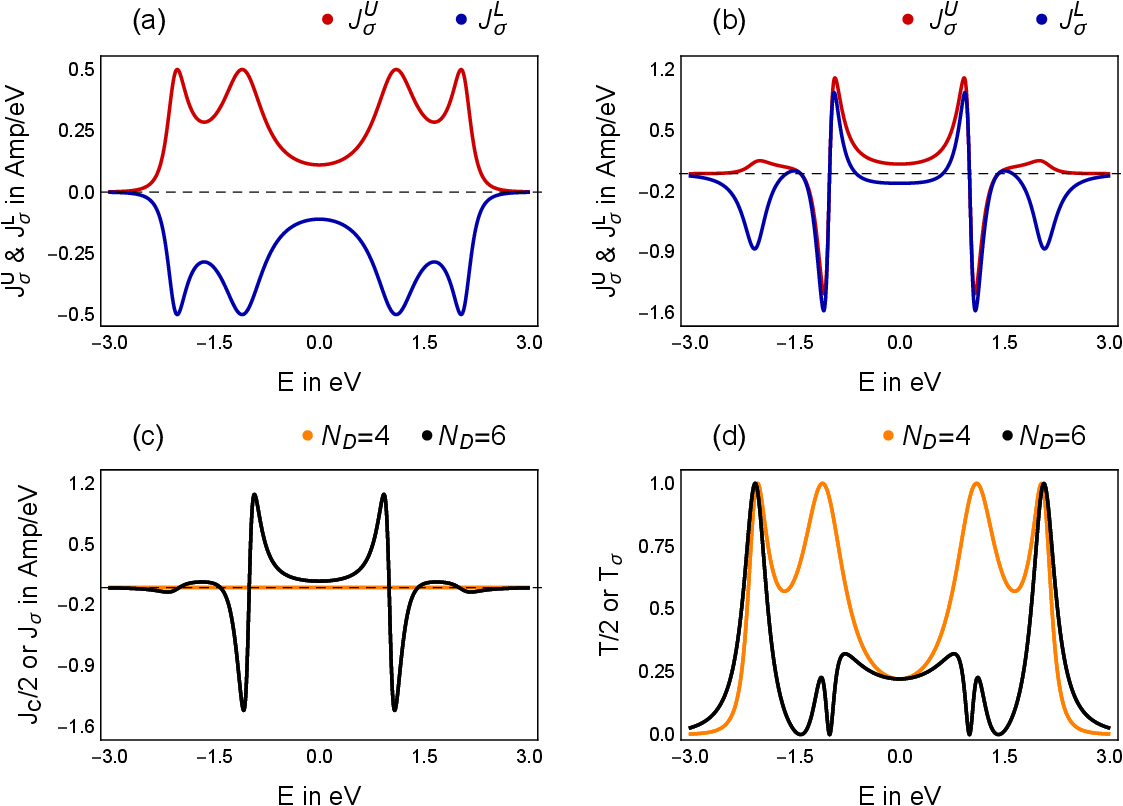}}\par}
\caption{(Color online). Current densities in the upper and lower arms for (a) symmetric and (b) most asymmetric-rings.
(c) Net circular current density and (d) transmission function for symmetric and most asymmetric-rings.
The ring has zero SOI and zero magnetic flux.}
\label{f3}
\end{figure}
(where the drain is attached) is called upper arm and the path between $N_D$ to 1 is called
lower arm. In a symmetric-ring, the source and drain are connected to the ring in such a way that,
the lengths of upper and lower arms are the same. Otherwise it is asymmetric-ring. If we connect the drain
to the $N$-th site of the ring, the path difference between upper and lower arms becomes maximum. This is
called most asymmetric-ring. Due to the conservation of charge, the bond charge current (or the bond
charge current density) in each bond of a particular arm (upper or lower) of the ring always remains the same.
In the absence of any spin-scattering interaction, the contribution of up and down spin
to the current are equal and current due to spin-flipping scattering becomes zero. That is,
\begin{eqnarray}
J_{\uparrow\uparrow, n \rightarrow n+1}(E) & = & J_{\downarrow\downarrow, n \rightarrow n+1}(E); \nonumber \\
J_{\uparrow\downarrow, n \rightarrow n+1}(E) & = & J_{\downarrow\uparrow, n \rightarrow n+1}(E) = 0.
\label{eq0a}
\end{eqnarray}
\noindent Hence,
\begin{eqnarray}
J_{\uparrow, n \rightarrow n+1}(E) & = & J_{\downarrow, n \rightarrow n+1}(E).
\label{eq0b}
\end{eqnarray}
\noindent
Now, the current flowing through the each bond of upper are equal to each other. That is,
\begin{equation}
J_{\sigma, n \rightarrow n+1}^A(E) = J_{\sigma, n+1 \rightarrow n+2}^A(E).
\label{eq01}
\end{equation}
\noindent $A = U,L$.
As all the bonds in an arm carry equal current we simply denote the bond current density in
the upper and lower arm as $J_{\sigma}^{\mbox{\tiny U}}(E)$ and $J_{\sigma}^{\mbox{\tiny L}}(E)$, respectively.
$J_{\sigma}^U (E)$ (red curve) and $J_{\sigma}^L$ (blue curve)
for symmetric and the most asymmetric ring are shown in the Fig.~\ref{f3}(a) - (b),
respectively. The net circular current density (Fig.~\ref{f3}(c)) and the overall junction transmission
function (Fig.~\ref{f3}(d)) for both the connections are also calculated.
The current densities are symmetric around $E=0$ (see in Appendix~\ref{aa}).
The energies associated with the picks and the dips in the each spectra correspond to the
energy eigenvalues of the ring. The energy dispersion relation for the isolated ring is
\begin{equation}
E_{\sigma} = 2 t \mbox{cos}(ka) = 2 t \mbox{cos}\left[\frac{2\pi m}{N}\right].
\label{dis1}
\end{equation}
\noindent $k = 2 \pi m/N a$. The integer $m$ runs between $N/2 \leq m <N/2$. Therefore for $+k$ and $-k$
the system has the same energy. In general for a quantum ring with any $N$,
the energy states have two-fold degeneracy due to periodic boundary condition except at $m=0$ for odd $N$
and $m= -N/2, 0$ for even $N$. For example, in this article, as we choose $N = 6$, the allowed values of
$m$ are $-3,~-2,~-1,~0,~1,~2$. Therefore the eigenstates with $m = \pm -2$ and $\pm 1$. Whereas the eigenstates
with $m = -3$, that is $E = -2\,$eV and $m = 0$ with $E = 2\,$eV, are non-degenerate.
In a symmetric-ring, the currents in the upper and lower arms are equal and opposite
to each other (Fig.~\ref{f3}(a)). Therefore we have four peaks in $J_{\sigma}^U-E$ (or $J_{\sigma}^L-E$) spectra.
The circular current density in terms of $J^{\mbox{\tiny U}}_{\sigma}$ and
$J^{\mbox{\tiny L}}_{\sigma}(E)$ can be written as
\begin{equation}
J_{\sigma}(E) = f^{\mbox{\tiny U}}J^{\mbox{\tiny U}}_{\sigma}(E) +
f^{\mbox{\tiny L}}J^{\mbox{\tiny L}}_{\sigma}(E).
\label{eqR1}
\end{equation}
\noindent $f^{\mbox{\tiny U}} = (N_D - 1)/N$ and $f^{\mbox{\tiny L}} = (N - N_D + 1)/N$ are the weight
factors for the upper and lower arms, respectively.
As we have already discussed, in symmetric-ring,
\begin{equation}
J_{\sigma}^{\mbox{\tiny U}}(E) = - J_{\sigma}^{\mbox{\tiny L}}(E)~~~~\sigma = \uparrow,~\downarrow.
\label{eq03}
\end{equation}
\noindent Hence the net circular current is zero. 

Fig.~\ref{f3}(b), the degeneracy between clockwise and anti-clockwise moving electronic wave-functions
are lifted due to the asymmetric ring-to-leads connection. Hence the currents in the two arms become unequal.
The currents associated with the degenerate energy levels propagate in the same direction, which is very unconventional
as through one arm current propagates against the bias. For non-degenerate
energy levels they move in opposite directions similar to Fig.~\ref{f3}(a). A net circular current
within the ring is produced within the ring as
\begin{equation}
J_{\sigma}^{\mbox{\tiny U}}(E) \neq J_{\sigma}^{\mbox{\tiny L}}(E)~~~\sigma = \uparrow, \downarrow.
\label{eq02}
\end{equation}
\noindent As across $E = \pm 2\,$eV, the current propagates in opposite directions, through the two arms of the
ring, vanishingly small current densities are obtained according to Eq.~\ref{eqR1}.
Whereas at the degenerate energies (neglecting spin degeneracy),
the contributions from both of the arms are additive, hence a net circular current
density is obtained as we can see in the Fig.~\ref{f3}(c). As in this case $J_{\uparrow}(E) = J_{\downarrow}(E)$,
hence $J_C(E) = J_{\uparrow}(E) + J_{\downarrow}(E) = 2 J_{\uparrow}(E) = 2 J_{\downarrow}(E)$. 

In the transmission spectra (Fig.~\ref{f3}(d)), we have four pecks in the symmetric case where as
it has six peaks in the asymmetric case indicating removal of degeneracies same as previous results.
We have anti-resonant states with $T_{\sigma}(E) = 0$ in the asymmetric connection at the degeneracies $E = \pm 1\,$eV.
Here to note that, $T_{\sigma}(E)$ is positive with a upper bound 1, whereas $J_{\sigma}(E)$ can have positive and
negative values with no bound. 

\subsection{In the presence of magnetic-field}

We consider a net magnetic field passing through the center of the ring. The direction of the field is perpendicular
to the confining plane of ring. In the presence of magnetic-field, the Hamiltonian
\begin{figure}[ht]
{\centering \resizebox*{8cm}{6cm}{\includegraphics{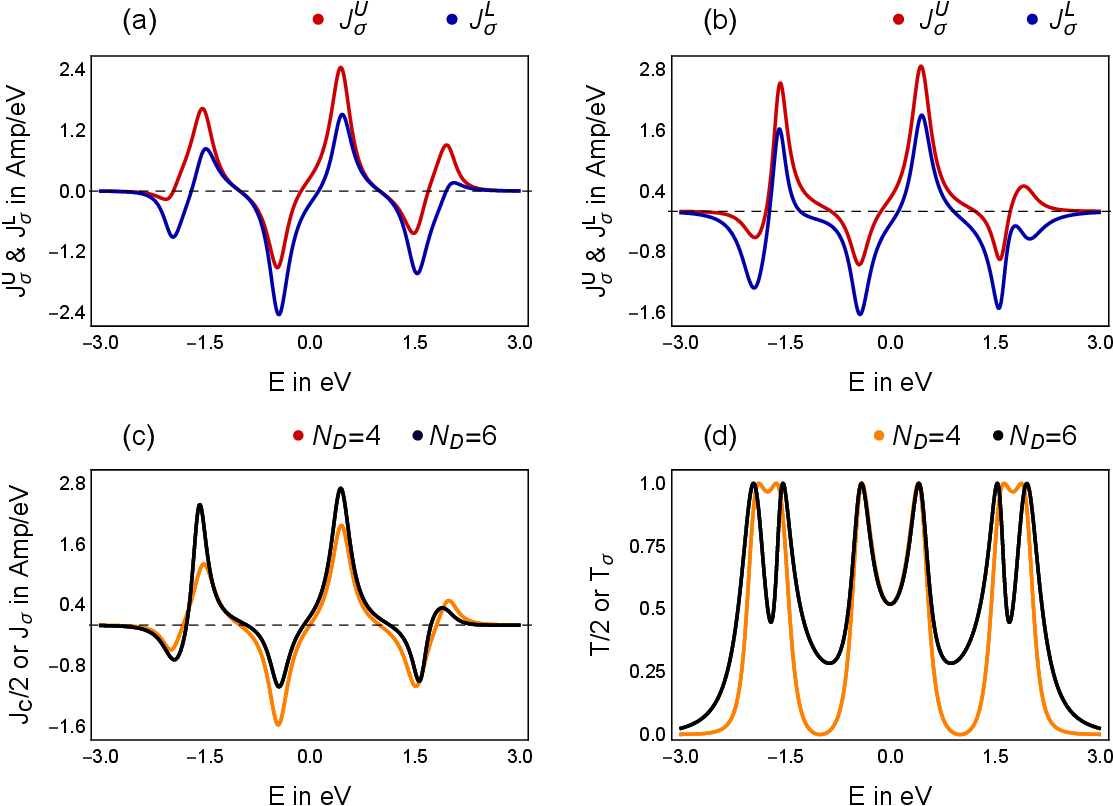}}\par}
\caption{(Color online). Same as Fig.~\ref{f3} in the presence of non-zero magnetic field ($\phi = 0.3$).}
\label{f4}
\end{figure}
$\mbox{\boldmath $H$}_{\mbox{\tiny Scatterer}}$ in Eq.~\ref{eq3} is modified as
\begin{equation}
\mbox{\boldmath $H$}_{\mbox{\tiny Scatterer}} = \sum_n \left(e^{i\theta}\mbox{\boldmath $c$}_{n+1}^{\dag} 
\mbox{\boldmath $t$} \mbox{\boldmath $c$}_n + 
e^{-i\theta}\mbox{\boldmath $c$}_{n}^{\dag} 
\mbox{\boldmath $t$}^{\dag} \mbox{\boldmath $c$}_{n+1} \right)
\label{eqMag1}
\end{equation}
where $\theta=2 \pi \phi/N$ is the phase factor due to the flux $\phi$
which is measured in unit of the elementary flux quantum $\phi_0$ ($=ch/e$),
and $n=1$, $2$, $3$ $\dots$. The circular current densities and the transmission function as a function of $E$
in the presence of magnetic field is shown in the Fig.~\ref{f4}.
Six peaks are visible in each spectrum as the Kramer's degeneracy gets removed with the in corporation of
magnetic-field. The energy dispersion relation is~\cite{gefen} 
\begin{equation}
E_{\uparrow} = E_{\downarrow} =
2 t \mbox{cos}\left[\frac{2\pi}{N}\left(m+\frac{\phi}{\phi_0}\right)\right].
\label{dis2}
\end{equation}
\noindent $m = 0,~\pm 1~\pm2 \dots$.
With $\phi = 0.3$, the eigen energies of a $6$-th sites ring are $E_{\sigma} = \pm 1.90,~ \pm 1.486, ~\pm 0.4158\,$eV.
Therefore energy levels are symmetric around $E = 0$. But the magnitudes of $J_{\sigma}^U (E)$ or $J_{\sigma}^L (E)$
are asymmetric around $E = 0$. Unlike Fig.~\ref{f3}(a), in the presence of non-zero magnetic field, in a symmetric-ring
$J_{\sigma}^U (E) \neq - J_{\sigma}^L (E)$ and they propagate in the same direction. Therefore net circular current
is non-zero in the symmetric-ring. In-fact a magnetic field can induces a net circulating current
with in a isolated ring (without any electrodes) and a lots of research have already been done in this
direction~\cite{ding,butt1,levy,jari,bir,chand,blu,ambe,schm1,schm2,peet,spl}. So here in a symmetric-ring,
magnetic field driven circular current appears.
In this case (Fig.~\ref{f4}(a)), $J_{\sigma}^U (E) = - J_{\sigma}^L (-E)$.
Therefore net $J_C (E)$ is anti-symmetric around $E = 0$ as we can see in the Fig.~\ref{f4}(c) shown by orange curve.
Therefore Eq.~\ref{j4} implies that the current goes to zero if we set the equilibrium Fermi energy $E_F$ to 0.
On the other hand under, in asymmetric-ring $J_C(E)$ is asymmetric around $E =0$ (shown by
the black curve in Fig.~\ref{f4}(c)). Therefore in this case $I_C(V)$ is finite for any
choice of $E_F$. The transmission spectra are symmetric (Fig.~\ref{f4}(d)) around $E = 0$.
Unlike Fig.~\ref{f3}(d), here the anti-resonances in this spectra get removed in an asymmetric-ring (as shown by
the black curve).

\subsection{In the presence of Rashba spin-orbit interaction}

We consider a ring scatterer with spin-orbit interaction. The rest parts of the circuit remains SOI free.
Here we take Rashba SOI (RSOI) which is originated due to the
structure inversion-asymmetry, caused by the inversion asymmetry of the confining
potential~\cite{Rashba0}. RSOI~\cite{Rashba1,Rashba2} is an electrically tunable spin-orbit
interaction~\cite{RashbaTune}.
In the presence of RSOI, the Hamiltonian $\mbox{\boldmath $H$}_{\mbox{\tiny Scatterer}}$ in Eq.~\ref{eq3} is modified as
\begin{eqnarray}
\mbox{\boldmath $H$}_{\mbox{\tiny Scatterer}} & = & \sum_n \left(\mbox{\boldmath $c$}_{n+1}^{\dag} 
\mbox{\boldmath $t$} \mbox{\boldmath $c$}_n + \mbox{\boldmath $c$}_{n}^{\dag} 
\mbox{\boldmath $t$}^{\dag} \mbox{\boldmath $c$}_{n+1} \right) \nonumber \\
& + &\sum\limits_{n = 1}^{N}
\left(\mbox{\boldmath $c$}_{n+1}^\dagger\left(i\mbox{\boldmath $\sigma$}_x\right)
\mbox{\boldmath $\alpha$}\cos\phi_{n,n+1}\mbox{\boldmath $c$}_n + h.c.\right)\nonumber \\
& - & \sum\limits_{n = 3}^{N_R + 2}\left(\mbox{\boldmath $c$}_{n+1}^\dagger\left(i\mbox{\boldmath $\sigma$}_y\right)
\mbox{\boldmath $\alpha$} \sin \phi_{n,n+1} \mbox{\boldmath $c$}_n + h.c.\right)\nonumber \\
& = & \sum\limits_{n = 3}^{N_R + 2}\left(\mbox{\boldmath $c$}_{n+1}^\dagger
\mbox{\boldmath $t$}_{\sigma\sigma',n \rightarrow n+1}^{\mbox{\tiny eff}} \mbox{\boldmath $c$}_n + \mbox{h.c.} \right).
\label{eq9}
\end{eqnarray}
\noindent
$\alpha$ is the strength of SOI. $\phi_{n \rightarrow n+1} = \left(\phi_n+\phi_{n+1}\right)/2,$ with
$\phi_n=2\pi (n-1)/N$ is the geometrical phase. $\mbox{\boldmath $\sigma$}_i$'s ($i=x$, $y$, $z$) are the Pauli spin
matrices in $\mbox{\boldmath $\sigma$}_z$ diagonal representation. The hopping operator
$\mbox{\boldmath $t$}_{\sigma\sigma', n\rightarrow n+1}^{\mbox{\tiny eff}}$ has the form
\begin{eqnarray}
\mbox{\boldmath $t$}_{\sigma\sigma', n\rightarrow n+1}^{\mbox{\tiny eff}} = \left(\begin{array}{cc}
t & i \alpha e^{-i\phi_{n \rightarrow n+1}}\\
i\alpha e^{i \phi_{n \rightarrow n+1}} & t\end{array}\right).
\label{hopSOI}
\end{eqnarray}
In terms of Green's functions the current density between any two successive bond can be written as
\begin{eqnarray}
\mbox{\boldmath $J$}_{\sigma\sigma', n\rightarrow n+1} & = &  \frac{1}{\hbar}
\left(\mbox{\boldmath $G$}^{C*}_{n_{\sigma}, n+1_{\sigma'}}\mbox{\boldmath $t$}^*_{\sigma\sigma'n \rightarrow n+1}
- h.c.\right).\nonumber \\
& &\label{eq10}
\end{eqnarray}
\noindent $\mbox{\boldmath $G$}^{C*}_{n_{\sigma}, n+1_{\sigma'}}$ is the correlated Green's function~\cite{coG1,coG2}.
In the presence of the SOI, the expressions of eigenvalues of a quantum ring are~\cite{maiti}:
\begin{eqnarray}
E_{\uparrow} & = & -2 t\mbox{cos}(\pi/N)\mbox{cos}(ka+\pi/N)\nonumber \\
& + & 2\mbox{sin}(ka+\pi/N)\sqrt{t^2\mbox{sin}^2(\pi/N)+\alpha^2}
\label{ras1}
\end{eqnarray}
\noindent and
\begin{eqnarray}
E_{\downarrow} & = & -2 t\mbox{cos}(\pi/N)\mbox{cos}(ka+\pi/N)\nonumber \\
& - & 2\mbox{sin}(ka+\pi/N)\sqrt{t^2\mbox{sin}^2(\pi/N)+\alpha^2}.
\label{ras2}
\end{eqnarray}
For example with $\alpha = 0.2\,$eV, our the eigen values of our present ring scattering
are $\pm 2.03852,~ \pm 2.03852,~ \pm 1.07703,~ \pm 1.07703,~ \pm 0.961484,~ \pm0.961484\,$eV.
Therefore they are symmetric around $E = 0$.

The Rashba spin-orbit interaction causes a momentum-dependent spin splitting of electronic bands.
\begin{figure}[ht]
{\centering \resizebox*{6cm}{4cm}{\includegraphics{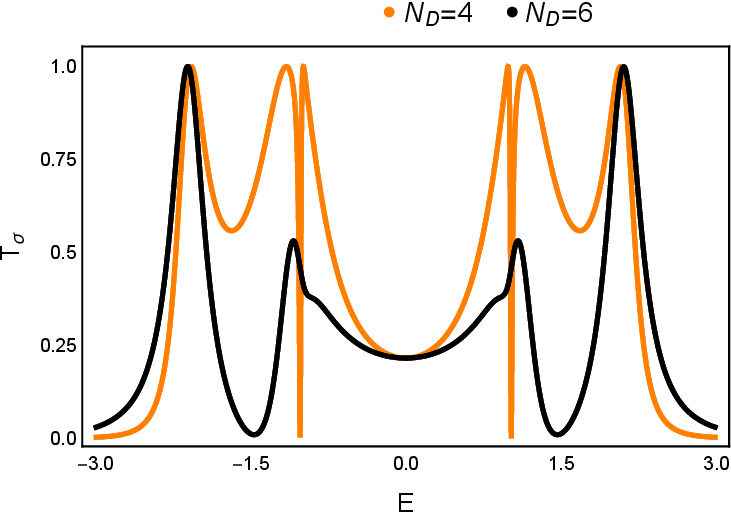}}\par}
\caption{(Color online). Transmission function vs energy for two different electrodes-to-ring configurations
in the presence of RSOI ($\alpha = 0.2\,$eV).}
\label{f5}
\end{figure}
As the time reversal symmetry remains preserve in a two terminal SOI device, we do not observe any
spin-separation in the overall junction current for a symmetric as well as asymmetric-rings.
The transmission function is plotted in Fig.~\ref{f5}.
\begin{figure*}
{\centering \resizebox*{16cm}{14cm}{\includegraphics{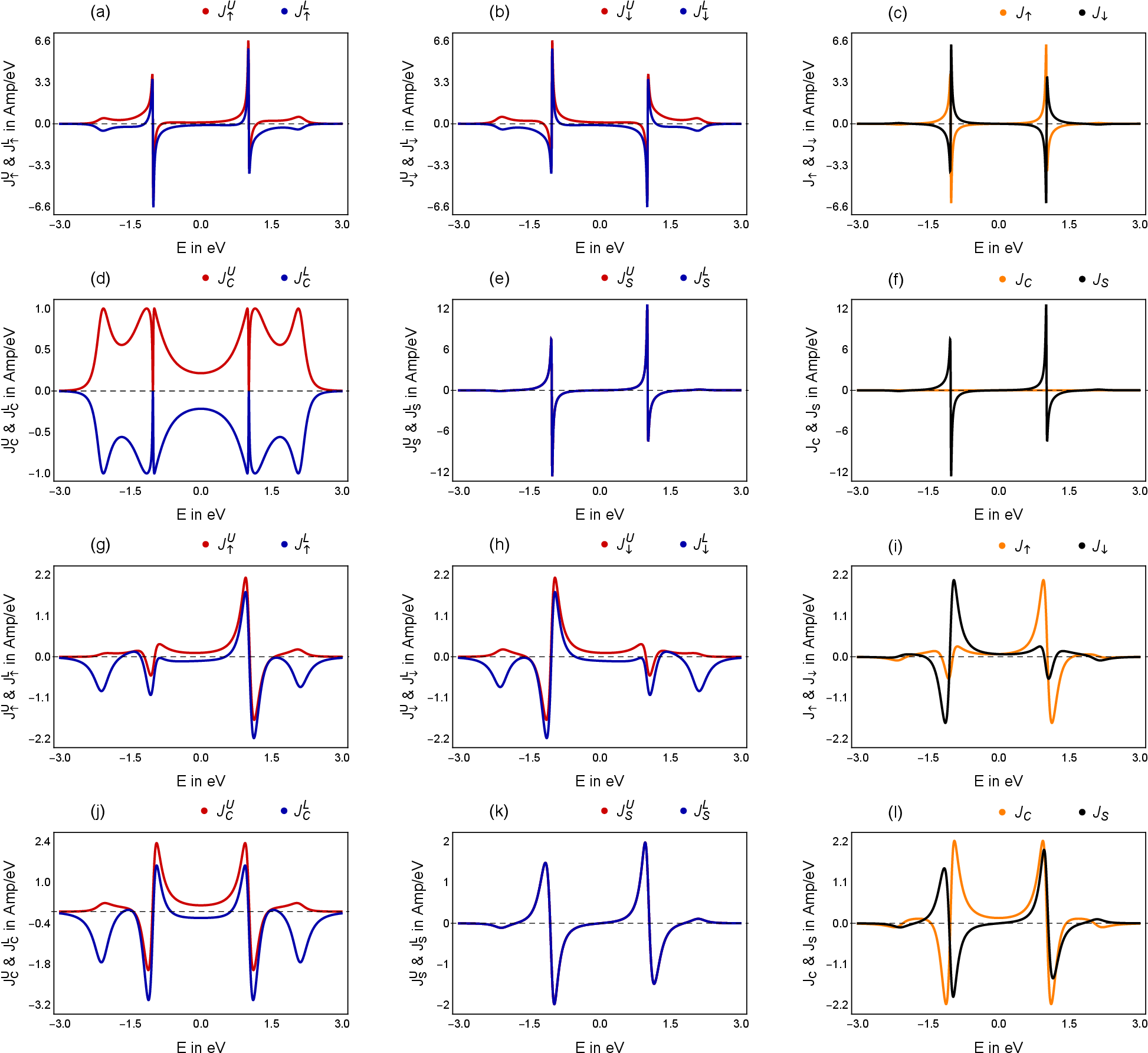}}\par}
\caption{(Color online). Circular current densities with energy for (a) - (f)
symmetric and (g) - (l) most asymmetric-rings in the presence of RSOI ($\alpha = 0.2\,$eV). Here we show the variation
of the following quantities with energy; (a) and (g) - Circular up spin
current density in the upper and lower arms of the ring; (b) and (h) Circular down spin current density in the upper and
lower arms of the ring; (c) and (i) Net up and down circular current densities; (d) and (j) Circular charge
current density in the upper and lower arms of the ring; (e) and (k) Circular spin
current density in the upper and lower arms of the ring; (f) and (l) Net circular charge and spin currents densities.}
\label{f6}
\end{figure*}
But within the channel a net spin current appears in both cases.
The circular current densities are plotted in Fig.~\ref{f6}.
Below, we summarize the relationship between spin-dependent current densities in the presence
of RSOI.
\vskip 0.2cm
\noindent
$\bullet$ Case I - Symmetric-ring with RSOI:

\begin{enumerate}

\item[a.] No spin-splitting is seen in the transmission function, i.e., $T_{\uparrow\downarrow}(E) =
T_{\downarrow\uparrow}(E) = 0$ and $T_{\uparrow\uparrow}(E) = T_{\downarrow\downarrow}(E)$ or
$T_{\uparrow}(E) = T_{\downarrow}(E)$. Transmission function is symmetric arrow $E=0$ i.e.,
$T_{\sigma}(E) = T_{\sigma}(-E)$; $\sigma = \uparrow,~\downarrow$ as shown in Fig.~\ref{f5}.

\item[b.] In the presence of spin-flip scattering, different
components of circular current do not remain conserved for each bond of the upper/lower arm. That is,
$J_{\sigma \sigma, n \rightarrow n+1}^A(E) \neq J_{\sigma \sigma, n+1 \rightarrow n+2}^A(E)$ and
$J_{\sigma \sigma', n \rightarrow n+1}^A(E) \neq J_{\sigma \sigma', n+1 \rightarrow n+2}^A(E)$.
$\sigma = \uparrow, \downarrow$, $\sigma' = \uparrow, \downarrow$ and $A = U, L$.

\item[c.] $J_{\sigma \sigma, n \rightarrow n+1}(E)$s are not symmetric around $E=0$ but
$J_{\sigma \sigma', n \rightarrow n+1}(E)$s are, i.e., $J_{\sigma \sigma, n \rightarrow n+1}(E)
\neq J_{\sigma \sigma, n \rightarrow n+1}(-E)$ but, $J_{\sigma \sigma', n \rightarrow n+1}(E)
= J_{\sigma' \sigma, n \rightarrow n+1}(-E)$. $\sigma = \uparrow, \downarrow$, $\sigma' = \uparrow, \downarrow$.

\item[d.] For a particular bond, $J_{\sigma \sigma, n \rightarrow n+1}(E) =
J_{\sigma' \sigma', n \rightarrow n+1}(-E)$ and $J_{\sigma \sigma', n \rightarrow n+1}(E) =
J_{\sigma' \sigma, n \rightarrow n+1}(E)$. $\sigma = \uparrow, \downarrow$, $\sigma' = \uparrow, \downarrow$.

\item[e.] Net up and down circular bond currents remains conserved in the upper/lower arm i.e.,
$J_{\sigma, n \rightarrow n+1}^A(E) = J_{\sigma, n+1 \rightarrow n+2}^A(E)$, $A = U, L$. Now onward we
simply denote them as $J_{\sigma}^U(E)$ or $J_{\sigma}^L(E)$ for the currents in the upper and lower arms,
respectively. They are asymmetric arrow $E=0$.

\item[f.] As shown in Fig.~\ref{f6} - (a), $J_{\uparrow}^U(E) = -J_{\uparrow}^L(-E)$.
Similarly in Fig.~\ref{f6} - (b) we find that $J_{\downarrow}^U(E) = -J_{\downarrow}^L(-E)$.
From both these figures i.e., Fig.~\ref{f6} - (a) - (b) we find that,
$J_{\uparrow}^U(E) = J_{\downarrow}^U(-E)$ and $J_{\uparrow}^L(E) = J_{\downarrow}^L(-E)$.
The net circular up and down spin current densities are plotted in Fig.~\ref{f6}(c). Here
$J_{\uparrow} (E) = - J_{\uparrow} (-E)$ and $J_{\downarrow} (E) = - J_{\downarrow} (-E)$.
And $J_{\uparrow} (E) = J_{\downarrow} (-E)$. Both $J_{\uparrow} (E)$ and $J_{\downarrow} (E)$
are anti-symmetric around $E = 0$.

\item[g.] The total charge current flowing through each bond remains conserved even in the presence of SOI i.e.,
$J_{C, n \rightarrow n+1}^A(E) = J_{C, n+1 \rightarrow n+2}^A(E)$, $A = U, L$.
But they are equal and opposite in the upper and lower arms of the ring, such that
$J_{C}^U(E) = - J_{C}^L(E)$ as we can see in Fig.~\ref{f6}(d).

\item[h.] Net circular charge current is zero (Fig.~ref{f6}(f) orange line).

\item[i.] Spin circular bond current within the ring conductor remains conserved.
Even at the upper arm of the ring, the spin current is identical to that at the lower arm.
Due to their overlap, we only see one curve in Fig.~\ref{f6}(e). Net spin current is anti-symmetric around
$E =0$, i.e., $J_S(E) = -J_S(-E)$ (Fig.~\ref{f6}(f)).

\end{enumerate}

\vskip 0.2cm
\noindent
$\bullet$ Case II - Asymmetric ring with RSOI:

\begin{enumerate}

\item[a.] The transmission spectra have same feature as case I-(a).
See Fig.~\ref{f5} black curve.

\item[b.] Same as case I-(b).

\item[c.] Same as case I-(c).

\item[d.] Same as case I-(d).

\item[e.] Same as case I-(e).

\item[f.] Unlike case I-(f), in asymmetric-ring $J_{\uparrow}^U \neq -J_{\uparrow}^L(-E)$
(Fig.~\ref{f6} (g)) and $J_{\downarrow}^U \neq -J_{\downarrow}^L(-E)$ (Fig.~\ref{f6} (h).
However same as case I-(f), here we have $J_{\uparrow}^U(E) = J_{\downarrow}^U(-E)$ and
$J_{\uparrow}^L(E) = J_{\downarrow}^L(-E)$ (as shown in Fig.~\ref{f6} (g) and (h)). 
$J_{\uparrow} (E) \neq - J_{\uparrow} (-E)$ and $J_{\downarrow} (E) \neq - J_{\downarrow} (-E)$
in this case (Fig.~\ref{f6} (i)). But $J_{\uparrow} (E) = J_{\downarrow} (-E)$ and vice-versa.
$J_{\uparrow} (E)$ and $J_{\downarrow} (E)$ are asymmetric around $E = 0$.

\item[g.] Same as symmetric-ring, the total charge current is also conserved here. In an asymmetric-ring,
$J_C^U(E)$ is not equal and opposite to $J_C^L(E)$ ( Fig.~\ref{f6}(j)).

\item[h.] Unlike symmetric-ring, net circular charge current is not zero here (Fig.~\ref{f6}(l) orange line). $J_C(E)$
is symmetric around $E = 0$.

\item[i.] The spin current densities in the upper and lower arms are plotted in Fig~\ref{f6}(k) and the
net spin current density is shown in Fig.~\ref{f6}(l) by black line. Their characteristics features are as in case I-(i).

\end{enumerate}

Therefore, a pure spin current appears (charge current is zero) in a symmetric ring.
As the spin current density is anti-symmetric around $E=0$, the net spin current $I_S$ (given
by the area under the $J_S - E$ curve (Eq.~(\ref{j4}))) is zero under the condition $E_F = 0$.
Therefore to get pure spin current, we need to set Fermi energy $E_F$ other than zero.
In an asymmetric-ring, we have a net charge as well as spin circular currents. As in this case,
$J_C(E)$ is symmetric around $E = 0$ and $J_S(E)$ is anti-symmetric around $E=0$, therefore at $E_F =0$,
the spin $I_S(V)$ current vanishes and we can get a pure charge current. Therefore we can achieve a pure
spin-to-charge conversion and vice-versa by selectively choosing the equilibrium  Fermi-energy and
the position of the drain. The phenomena can be explained as follows.

In the presence of RSOI, the up and down spins move in the opposite directions.
When an electron with charge $e$ and momentum $\bf{p}$  moves in a magnetic field $\bf{B}$,
Lorentz force ${\bf F}=-e({\bf p}\times {\bf B})/m$ is acted on it in the direction perpendicular to its motion.
Similarly, when an electron moves in an electric field $\mbox{\boldmath $\xi$} = \mbox {\boldmath $\nabla$} V$,
it experiences a magnetic field ${\bf B}_{eff} \sim \mbox{\boldmath $\xi$} \times {\bf p}/mc^2$
in its rest-frame~\cite{Rashba2}. In quantum wells with broken structural inversion
symmetry, the inter-facial electric field $\mbox{\boldmath $\xi$}$ along the $z$-direction gives rise to the
RSOI coupling $(\mbox{\boldmath $\sigma$} \times {\bf p})_z = \sigma_x p_y - \sigma_y p_x$. Therefore
when a spin moves in the $x-y$ plane, it experiences a spin-dependent Lorentz force due to the effective
magnetic field ${\bf B}_{eff}$~\cite{lorentz1,lorentz2,mp16}. This force is proportional to the square of the
transverse electric field $\xi$.
Here we consider a ring geometry that has doubly degenerate energy states, representing electronic wave functions
with equal and opposite momenta. For symmetric-ring, when these degeneracies remain preserved, the velocities
of the up and down spin are exactly equal and opposite to each
other. For asymmetric-ring, the degeneracy gets lifted. As a result, the velocities of the spins are not equal but
opposite to each other.

\subsection{Effect of magnetic field and RSOI}

In the presence of both SOI and magnetic field the Hamiltonian $\mbox{\boldmath $H$}_{\mbox{\tiny Scatterer}}$
in Eq.~\ref{eq3} is modified as:
\begin{eqnarray}
\mbox{\boldmath $H$}_{\mbox{\tiny Scatterer}} & = & \sum_n \left(e^{i\theta}\mbox{\boldmath $c$}_{n+1}^{\dag} 
\mbox{\boldmath $t$} \mbox{\boldmath $c$}_n +  e^{-i\theta}\mbox{\boldmath $c$}_{n}^{\dag} 
\mbox{\boldmath $t$}^{\dag} \mbox{\boldmath $c$}_{n+1} \right) \nonumber \\
& - & \sum_n\alpha\left[\mbox{\boldmath $c$}_{n+1}^{\dag} 
\left(i\mbox{\boldmath $\sigma$}_x \cos\varphi_{n,n+1} + 
i\mbox{\boldmath $\sigma$}_y \sin\varphi_{n,n+1}\right)\right.\nonumber \\
& & \left. e^{i\theta} 
\mbox{\boldmath $c$}_n + h.c. \right]
\label{hamiMagSOI}
\end{eqnarray}
\noindent In the presence of SOI and magnetic field, the hopping operator
$\mbox{\boldmath $t$}_{\sigma\sigma', n \rightarrow n+1}$ (Eq.~\ref{hopSOI}) is modified as
\begin{eqnarray}
\mbox{\boldmath $t$}_{\sigma\sigma', n \rightarrow n+1} = e^{-i\theta}\left(\begin{array}{cc}
t & i \alpha e^{-i\phi_{n \rightarrow n+1}}\\
i\alpha e^{i \phi_{n \rightarrow n+1}} & t\end{array}\right).\nonumber \\
\label{hopSOIMag}
\end{eqnarray}
\noindent The expression for the energy eigen values are~\cite{maiti}:
\begin{eqnarray}
E_{\uparrow} & = & -2 t\mbox{cos}(\pi/N)\mbox{cos}(ka+\theta+\pi/N)\nonumber \\
& + & 2\mbox{sin}(ka+\theta+\pi/N)\sqrt{t^2\mbox{sin}^2(\pi/N)+\alpha^2}
\label{magras1}
\end{eqnarray}
\noindent and
\begin{eqnarray}
E_{\downarrow} & = & -2 t\mbox{cos}(\pi/N)\mbox{cos}(ka+\theta+\pi/N)\nonumber \\
& - & 2\mbox{sin}(ka+\theta+\pi/N)\sqrt{t^2\mbox{sin}^2(\pi/N)+\alpha^2}
\label{magras2}
\end{eqnarray}
\noindent Therefore in the presence of magnetic field and SOI, the spin and Kramer's degeneracies are
\begin{figure}
{\centering \resizebox*{8cm}{5.5cm}{\includegraphics{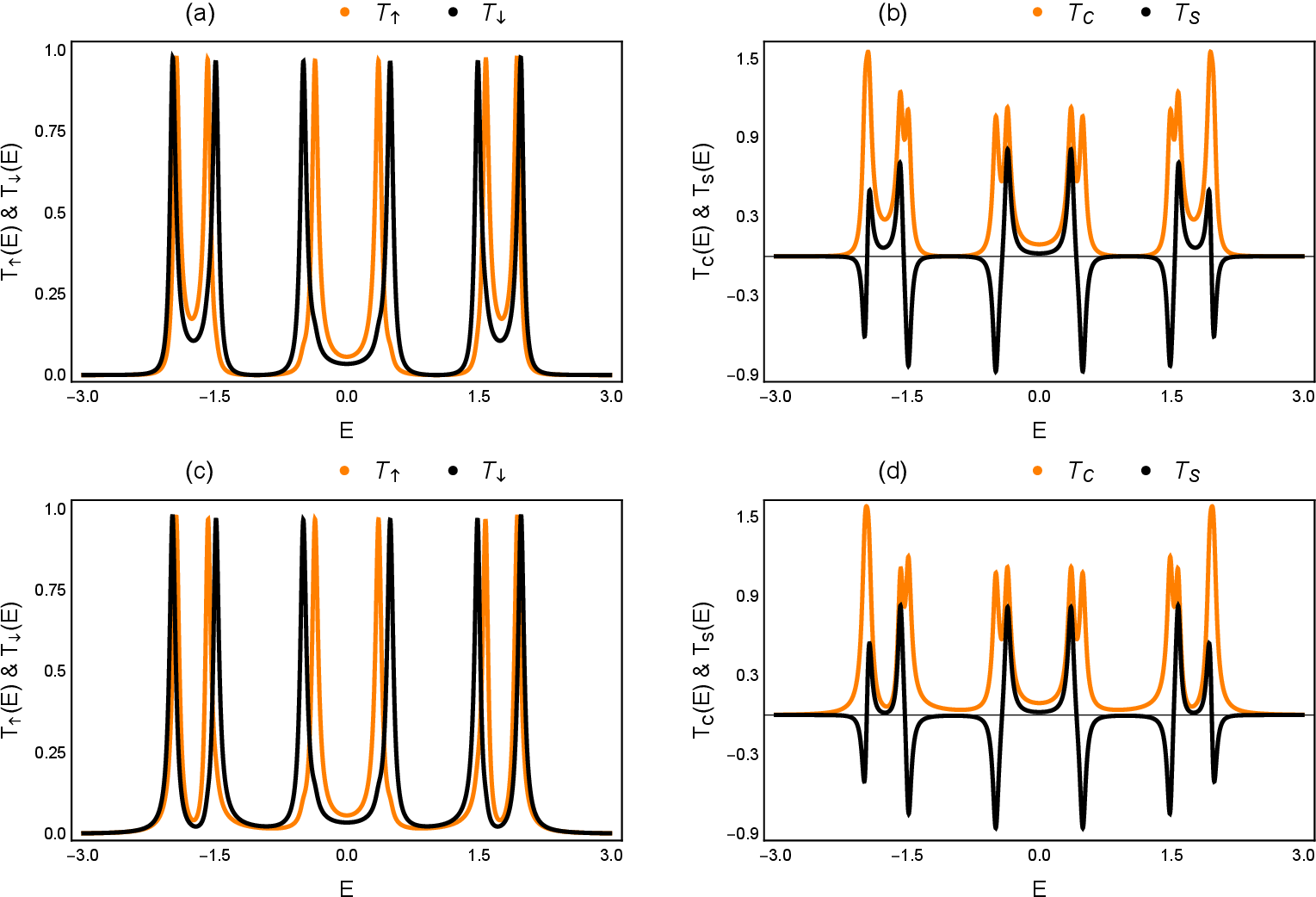}}\par}
\caption{(Color online). Transmission functions as a function of energy for (a) - (b) symmetric
and (c) - (d) most asymmetric-rings in the presence of magnetic field ($\phi = 0.3$) and RSOI ($\alpha = 0.2\,$eV).
In the figures (a) and (c) we plot the net up and down transmission functions and in the figures (b) and (d)
we show net charge and spin transmission functions.}
\label{f7}
\end{figure}
completely removed. For example, with $\phi = 0.3$ and $\alpha = 0.2\,$eV a $6-$th site ring has
$E = \pm 0.358577,~\pm 0.489086,~\pm 1.47027k,~\pm 1.55955,~\pm 1.91813,~\pm 1.95936\,$eV. So
the energy levels are symmetric around $E = 0$. Now we summarize the relationship between spin-dependent
\begin{figure*}
{\centering \resizebox*{16cm}{14cm}{\includegraphics{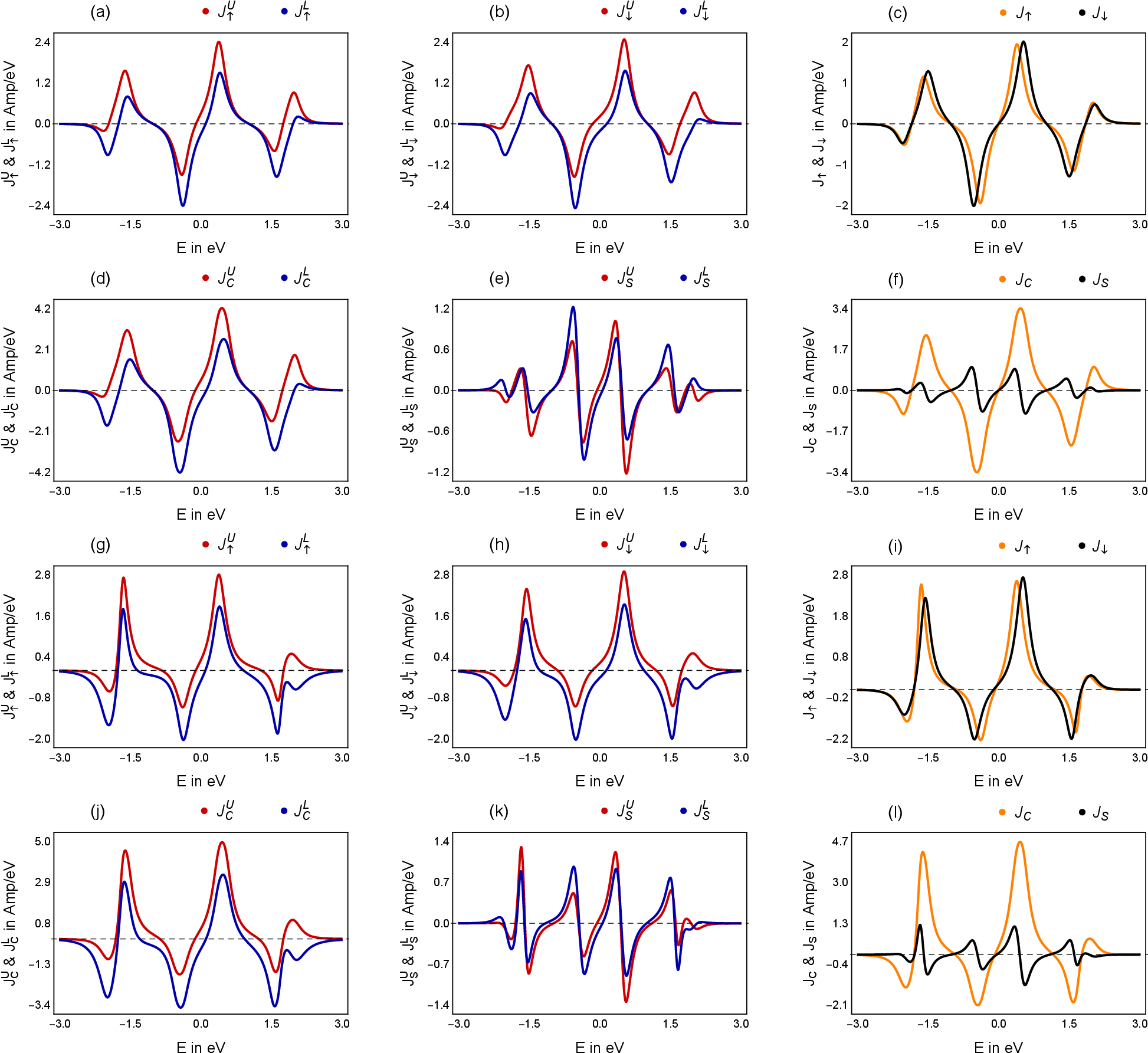}}\par}
\caption{(Color online). Circular current densities as a function of energy
in the presence of RSOI ($\alpha = 0.2\,$eV) and magnetic field ($\phi = 0.3$). The meaning of the figures
are same as in Fig.~\ref{f6}.}
\label{f8}
\end{figure*}
current densities in the presence of magnetic field and spin-orbit interaction.
\vskip 0.2cm
\noindent
$\bullet$ Case III - Symmetric-ring with magnetic field and RSOI:

\begin{enumerate}

\item[a.] Spin separation is now seen in the transmission function where,
$T_{\uparrow\uparrow}(E) \neq T_{\downarrow\downarrow}(E)$. The spin-flipping parts are identical
i.e., $T_{\uparrow\downarrow}(E) = T_{\downarrow\uparrow}(E) \neq 0$. Therefore net
up  $T_{\uparrow }(E)( = T_{\uparrow\uparrow}(E) + T_{\downarrow\uparrow}(E))$
and down $T_{\downarrow }(E)( = T_{\downarrow\downarrow}(E) + T_{\uparrow\downarrow}(E))$
are not the same (see Fig.~\ref{f7}(a)). Hence a net spin transmission $T_S(E) = T_{\uparrow }(E) - T_{\downarrow }(E)$
is seen along with charge transmission i.e, $T_C(E) = T_{\uparrow }(E) + T_{\downarrow}(E)$ (see Fig.~\ref{f7}(b)).
All these transmission functions are symmetric arrow $E=0$ i.e., $T_{\sigma\sigma'}(E) = T_{\sigma\sigma'}(-E)$.
$\sigma = \uparrow,~\downarrow$ and $\sigma' = \uparrow,~\downarrow$. Therefore $T_{\uparrow} (E)$,
$T_{\downarrow }(E)$, $T_S(E)$, and $T_C(E)$ are also symmetric around $E = 0$.

\item[b.] Same as Case I-(b).

\item[c.] All the spin components of the circular bond current densities are asymmetric around $E=0$.

\item[d.] For a particular bond, unlike case I-(d), $J_{\uparrow\uparrow, n \rightarrow n+1}(E) =
J_{\downarrow\downarrow, n \rightarrow n+1}(-E)$, but similar to  case I-(d),
$J_{\uparrow\downarrow, n \rightarrow n+1}(E) = J_{\downarrow\uparrow, n \rightarrow n+1}(E)$.

\item[e.] Same as Case I-(e).

\item[f.] Similar to Case I-(f), $J_{\uparrow}^U(E) = - J_{\uparrow}^L(-E)$ (Fig.~\ref{f8}(a))
and $J_{\downarrow}^U(E) = - J_{\downarrow}^L(-E)$ (Fig.~\ref{f8}(b)). But here,
$J_{\uparrow}^A(E) \neq J_{\downarrow}^A(-E)$, and vice-versa. $A = U$ or $L$. Net
circular up and down spin currents are anti-symmetric around $E = 0$, i.e., 
$J_{\uparrow} (E) = - J_{\uparrow} (-E)$ and $J_{\downarrow} (E) = - J_{\downarrow} (-E)$ (Fig.~\ref{f8}(c)).
Unlike, Case I-(f), here we see $J_{\uparrow} (E) \neq J_{\downarrow} (-E)$.

\item[g.] The total charge current flowing through each bond remains conserved. $J_C (E)$ in the upper and lower arms
are not equal. They propagate in the same direction around each eigen energy. $J_C^U(E) = -J_C^L(-E)$ (Fig.~\ref{f8}(d)).

\item[h.] $J_C(E)$ is anti-symmetric around $E=0$ (Fig.~\ref{f8}(f) shown by orange curve).

\item[i.] Circular bond spin current within the ring conductor remains conserved, though
$J_S^U(E)$ and $J_S^L(E)$ are not equal. $J_S^U(E) = -J_S^L(-E)$ (Fig.~\ref{f8}(e)). $J_S(E)$ is anti-symmetric
around $E=0$ (Fig.~\ref{f8}(f) shown by black curve).

\end{enumerate}

$\bullet$ Case IV - Asymmetric-ring with magnetic field and RSOI:

\begin{enumerate}

\item[a.] Similar to symmetric-ring, $T_{\uparrow\uparrow}(E) \neq T_{\downarrow\downarrow}(E)$ and
$T_{\uparrow\downarrow}(E) = T_{\downarrow\uparrow}(E) \neq 0$. The net
up  $T_{\uparrow }(E)$ and down $T_{\downarrow }(E)$ transmission are also not the same (see Fig.~\ref{f7}(c)).
Hence $T_S(E)$ and $T_C(E)$ are both non-zero here (see Fig.~\ref{f7}(d)). But unlike symmetric case,
$T_{\uparrow\uparrow}(E),~T_{\uparrow\downarrow}(E),~T_{\downarrow\downarrow}(E),~T_{\downarrow\uparrow}(E)$,
are asymmetric around $E=0$. But $T_{\uparrow}(E),~T_{\downarrow }(E), T_C(E)$ and $T_S(E)$ are
symmetric around $E = 0$.

\item[b.] Same as case I-(b).

\item[c.] $J_{\uparrow\uparrow}(E)$, $J_{\uparrow\downarrow}(E)$, $J_{\downarrow\uparrow}(E)$ and
$J_{\downarrow\downarrow}(E)$ are asymmetric around $E = 0$.

\item[d.] Unlike case I-(b), for a particular, $J_{\sigma \sigma, n \rightarrow n+1}(E) \neq
J_{\sigma' \sigma', n \rightarrow n+1}(-E)$. But $J_{\sigma \sigma', n \rightarrow n+1}(E) =
J_{\sigma' \sigma, n \rightarrow n+1}(E)$. $\sigma = \uparrow, \downarrow$, $\sigma' = \uparrow, \downarrow$.

\item[e.] Same as case I-(e).

\item[f.] Unlike case I-(f), $J_{\uparrow}^U(E) \neq - J_{\uparrow}^L(-E)$ (Fig.~\ref{f8}(g)) and
$J_{\downarrow}^U(E) \neq - J_{\downarrow}^L(-E)$ (Fig.~\ref{f8}(h)). $J_{\uparrow}^A(E) \neq J_{\downarrow}^A(-E)$
and vice-versa ($A = U$ or $L$). Net $J_{\uparrow} (E)$ and $J_{\downarrow} (E)$ are asymmetric around $E = 0$
(Fig.~\ref{f8}(i)). $J_{\uparrow} (E)$ is not equal to $J_{\downarrow} (-E)$ and vice-versa.

\item[g.] The charge current is also conserved here. In asymmetric-ring, $J_C^U(E)$ is not equal
and opposite to $J_C^L(E)$ (Fig.~\ref{f8}(j)). Rather they flow in the same direction in the two
arms for the most of the energy window.

\item[h.] The net circular charge current is non-zero (Fig.~\ref{f8}(l), shown by orange curve). $J_C(E)$ is
asymmetric around $E=0$.

\item[i.] Same as case I-(i) $J_S(E)$ is conserved. But unlike case I-(i) they are not the same in
the upper and lower arms (Fig.~\ref{f8}(k)). The net circular spin current density is asymmetric around $E=0$
(Fig.~\ref{f8}(l), shown by black curve).

\end{enumerate}

As in a symmetric ring $J_C(E)$ and $J_S(E)$ both are non-zero and anti-symmetric around $E = 0$,
thus circular charge and spin currents become zero if we set Fermi energy at 0. On the other hand
in an asymmetric-ring, these are asymmetric around $E = 0$. Therefore we have net circular charge and spin
currents for any choices of the Fermi energy.

\section{Conclusion}

Though the transmission function vs energy characteristics are always symmetric around $E = 0$,
for a perfect (without any disorder) open quantum ring, the circular current density - energy
spectra are affected by the degeneracy breaking. Our system is composed of a ring
attached with two-semi infinite electrodes. A quantum ring
has two-fold degeneracy as the electron moving in the clock and anti-clockwise
direction has the same energy. The corresponding degeneracy can be lifted by connecting the electrodes
asymmetrically to the ring. The time-reversal symmetry can be broken by adding magnetic-field to the
system. Whereas spin-degeneracy is lifted by adding spin-orbit interaction. The energy eigenvalues
of a perfect ring are symmetric around $E = 0$ in the presence or absence of
the magnetic field or SOI or both. The peaks in the current density and the transmission function
appear at these energy eigenvalues. The effects of these factors are summarized as follows:

\begin{enumerate}

\item[$\bullet$] In the absence of magnetic field and SOI, the circular charge current density $J_C(E)$
is zero in a symmetric-ring and symmetric around $E=0$ in an asymmetric-ring.
Thus net circular current is zero and non-zero for symmetric and asymmetric-rings, respectively.
Spin degeneracy is not broken here, thus spin current is zero.

\item[$\bullet$] In the presence of a magnetic-field, the charge circular current is anti-symmetric
and asymmetric around $E = 0$ for symmetric and asymmetric-rings, respectively. Thus in
symmetric-ring we have net non-zero circular charge current for $E_F \neq 0$.
For asymmetric-ring it is non-zero for any choice of $E_F$. The spin current is zero here.

\item[$\bullet$] In the presence of SOI, up and down spins separation happens within the ring. The circular charge
current density is zero for the symmetric-ring and it is non-zero and symmetric around $E = 0$ for the asymmetric-ring.
The circular spin current density is non-zero and anti-symmetric around $E = 0$ for both situations.
Therefore in a symmetric-ring we have pure spin circular current (as circular charge current is zero) setting the
Fermi-energy at any value other than zero. On the other hand, in an asymmetric-ring, we have both circular charge and
spin current for $E_F \neq 0$ and at $E_F = 0$ we have pure charge current (where spin current is zero).

\item[$\bullet$] In the presence of both magnetic field and SOI, in a symmetric-ring, both
the circular charge and spin current densities become non-zero. Both of them are anti-symmetric around $E = 0$.
Therefore we have non-zero $I_C$ and $I_S$ if $E_F \neq 0$, and at $E_F = 0$ both of them become zero.
For asymmetric-ring $J_C(E)$ and $J_S(E)$ are asymmetric around $E = 0$. Therefore we have both non-zero
circular charge and spin currents for any choices of the Fermi-energy.

\item[$\bullet$] The transmission function is always symmetric around $E = 0$. Therefore the junction
current is non-zero for any choices of $E_F$. Unless both the spin and Kramer's degeneracies are removed,
spin separation does not appear at the overall junction current. Therefore $T_S (E)$ is zero unless
both magnetic field and SOI become finite.

\end{enumerate}

As our study focuses on the qualitative nature of current inside and outside of a loop junction, The results
will serve to design the new generation electronic and spintronic devices.

\section{Acknowledgements}

The author acknowledges the financial support through the National Postdoctoral Fellowship (NPDF), SERB file
No. PDF/2022/001168.

\appendix

\section{Circular current density from group velocity}
\label{aa}

The sign and magnitude of current density at $\pm E$ can be explained from the group velocity
$v_k=\frac{1}{\hbar}\frac{\partial E}{\partial k}$. We can express the circular current density as
$J_C(E) \sim \frac{\partial E}{\partial k}$~\cite{fibocharge}. Therefore for a perfect ring,
the $J_C$ is proportional to $\sim -2 t \mbox{sin}(ka)$ (using Eq.~\ref{dis1}). Let us consider Fig.~\ref{f3}(a).
Here as we see, in the absence of any external interaction, for a symmetric-ring, the circular current
density in the upper or lower arm, is symmetric around $E=0$. Thus,
$J_U(E = 1\,\mbox{eV}) = J_U(E = -1\,\mbox{eV})$.
For an isolated ring, $E = 1\,$eV corresponds to $m = 2$ with $k = \frac{2\pi}{3}$ and $E = - 1\,$eV
corresponds to $m = 1$ with $k = \frac{\pi}{3}$ (here we consider only the positive values of the momentum).
Hence at $E = \pm 1\,$eV, $J_U (E) \sim 2 t \frac{\sqrt{3}}{2}$, hence $J_U(E)$ is symmetric around $E = 0$.

Similarly in the presence of interactions, from the dispersion relations stated in
Eq.~\ref{dis2}, Eq.~\ref{ras1}, Eq.~\ref{ras2}, Eq.~\ref{magras1}, and Eq.~\ref{magras2}, we can predict the
circular current at any $\pm E$.


\begin{thebibliography}{99}

\bibitem{cir1a} A. Nakanishi and M. Tsukada, Phys. Rev. Lett. \textbf{87},
126801 (2001).

\bibitem{cir2} K. Tagami and M. Tsukada, Curr. Appl. Phys. \textbf{3},
439 (2003).

\bibitem{cir3} M. Tsukada, K. Tagami, K. Hirose, and N. Kobayashi, J. Phys.
Soc. Jpn. \textbf{74}, 1079 (2005).

\bibitem{cir4} G. Stefanucci, E. Perfetto, S. Bellucci, and M. Cini,
Phys. Rev. B \textbf{79}, 073406 (2009).

\bibitem{Nitzan1} D. Rai, O. Hod, and A. Nitzan, J. Phys. Chem. C \textbf{114},
20583 (2010).

\bibitem{cir6} D. Rai, O. Hod, and A. Nitzan, Phys. Rev. B \textbf{85},
155440 (2012).

\bibitem{SKM} S. K. Maiti, Eur. Phys. J. B \textbf{86}, 296 (2013).

\bibitem{cir8} S. K. Maiti, J. Appl. Phys. \textbf{117}, 024306 (2015).

\bibitem{cir9} M. Patra and S. K. Maiti, Sci. Rep. \textbf{7}, 43343 (2017).

\bibitem{cir10} U. Dhakal and D. Rai, J. Phys.: Condens. Matter \textbf{31},
125302 (2019).

\bibitem{ring1} A. Lorke, R. J. Luyken, A. O. Govorov, J. P. Kotthaus, J. M. Garcia, and P. M. Petroff,
Phys. Rev. Lett. \textbf{84}, 2223 (2000).

\bibitem{ring2} R.J. Warburton, C. Sch\"{a}flein, D. Haft, F. Bickel, A. Lorke,
K. Karrai, J.M. Garcia, W. Schoenfeld, and P. M. Petroff, Nature (London) \textbf{405}, 926 (2000).

\bibitem{ring3} U. Fano, Phys. Rev. \textbf{124}, 1866 (1961).

\bibitem{ring4} J. G\"{o}res, et al., Phys. Rev. B \textbf{62} 2188 (2000).

\bibitem{mpprb} M. Patra and S. K. Maiti, Phys. Rev. B \textbf{100}, 165408 (2019).

\bibitem{jay1} A. M. Jayannavar and P. Singha Deo, Phys. Rev. B \textbf{51}, 10175 (1995).

\bibitem{jay2} S. Bandopadhyay, P. Singha Deo and A. M. Jayannavar, Phys. Rev. B \textbf{70}, 075315
(2004).

\bibitem{kra1} S. Lieu, M. McGinley, O. Shtanko, N. R. Cooper, and A. V. Gorshkov,
Phys. Rev. B \textbf{105}, L121104 (2022).

\bibitem{kra2} H. A. Kramers, Koninkl. Ned. Akad. Wetenschap., Proc. \textbf{33},
959 (1930).

\bibitem{kra3} E. Wigner, Nachrichten von der Gesellschaft der
Wissenschaften zu G\"{o}ttingen, Mathematisch-Physikalische
Klasse \textbf{1932}, 546 (1932).

\bibitem{spind} Lin, W., Li, L., Do\={g}an, F. et al. Nat Commun 10, 3052 (2019).

\bibitem{dev1} S. A. Wolf, D. D. Awschalom, R. A. Buhrman, J. M. Daughton,
S. von Moln\'{a}r, M. L. Roukes, A. Y. Chtchelkanova, and D. M. Treger,
Science \textbf{294}, 1488 (2001).

\bibitem{th1} I. Zutic, J. Fabian, and S. Das Sarma, Rev. Mod. Phys.
\textbf{76}, 323 (2004).

\bibitem{th2} S. Datta and B. Das, Appl. Phys. Lett. \textbf{56},
665 (1990).

\bibitem{th3} S. Bellucci and P. Onorato, Phys. Rev. B \textbf{78},
235312 (2008).

\bibitem{th4} P. F\"{o}ldi, B. Moln\'{a}r, M. G. Benedict, and F. M.
Peeters, Phys. Rev. B \textbf{71}, 033309 (2005).

\bibitem{th5} P. F\"{o}ldi, M. G. Benedict, O. K\'{a}lm\'{a}n, and F. M.
Peeters, Phys. Rev. B \textbf{80}, 165303 (2009).

\bibitem{th6} S. Bellucci and P. Onorato, J. Phys.: Condens. Matter
\textbf{19}, 395020 (2007).

\bibitem{th7} S.-Q. Shen, Z.-J. Li, and Z. Ma, Appl. Phys. Lett.
\textbf{84}, 996 (2004).

\bibitem{th8} D. Frustaglia and K. Richter, Phys. Rev. B \textbf{69},
235310 (2004).

\bibitem{th9} D. Frustaglia, M. Hentschel, and K. Richter, Phys.
Rev. Lett. \textbf{87}, 256602 (2001).

\bibitem{th10} M. Hentschel, H. Schomerus, D. Frustaglia, and K. Richter,
Phys. Rev. B \textbf{69}, 155326 (2004).

\bibitem{scatter1} V. B.-Moshe, D. Rai, S. S. Skourtis, and A. Nitzan, J. Chem. Phys. \textbf{133},
054105 (2010).

\bibitem{tb} J. C. Slater and G. F. Koster, Phys. Rev. \textbf{94}, 1498 (1954).


\bibitem{gefen} H.F. Cheung, Y. Gefen, E.K. Reidel, and W.H. Shih, Phys. Rev. B \textbf{37}, 6050
(1988).

\bibitem{ding} G. H. Ding and B. Dong, Phys. Rev. B \textbf{76}, 125301
(2007).

\bibitem{butt1} M. B\"{u}ttiker, Y. Imry, and R. Landauer, Phys. Lett. A
\textbf{96}, 365 (1983).

\bibitem{levy} L. P. L\'{e}vy, G. Dolan, J. Dunsmuir, and H. Bouchiat,
Phys. Rev. Lett. \textbf{64}, 2074 (1990).

\bibitem{jari} E. M. Q. Jariwala, P. Mohanty, M. B. Ketchen, and R. A.
Webb, Phys. Rev. Lett. \textbf{86}, 1594 (2001).

\bibitem{bir} N. O. Birge, Science \textbf{326}, 244 (2009).

\bibitem{chand} V. Chandrasekhar, R. A. Webb, M. J. Brady, M. B. Ketchen,
W. J. Gallagher, and A. Kleinsasser, Phys. Rev. Lett. \textbf{67}, 3578
(1991).

\bibitem{blu} H. Bluhm, N. C. Koshnick, J. A. Bert, M. E. Huber, and
K. A. Moler, Phys. Rev. Lett. \textbf{102}, 136802 (2009).

\bibitem{ambe} V. Ambegaokar and U. Eckern, Phys. Rev. Lett. \textbf{65},
381 (1990).

\bibitem{schm1} A. Schmid, Phys. Rev. Lett. \textbf{66}, 80 (1991).

\bibitem{schm2} U. Eckern and A. Schmid, Europhys. Lett. \textbf{18},
457 (1992).

\bibitem{peet} L. K. Castelano, G.-Q. Hai, B. Partoens, and F. M. Peeters,
Phys. Rev. B \textbf{78}, 195315 (2008).

\bibitem{spl} J. Splettstoesser, M. Governale, and U. Z\"{u}licke,
Phys. Rev. B \textbf{68}, 165341 (2003).

\bibitem{Rashba0} Y. Feng, et al., Nat Commun. \textbf{10}, 4765 (2019).

\bibitem{Rashba1} Y.A. Bychkov, E.I. Rashba, J. Exp. Theor. Phys. Lett.
\textbf{39}, 78 (1984)

\bibitem{Rashba2} A. Manchon, H. Koo, J. Nitta, S. M. Frolov, and R. A.
Duine, Nature Mater \textbf{14}, 871 (2015).

\bibitem{RashbaTune} L. Meier, G. Salis, I. Shorubalko, E. Gini,
S. Sch\"{o}n, K. Ensslin, Nat. Phys. \textbf{3}, 650 (2007).


\bibitem{coG1} L. Wang, K. Tagami, M. Tsukada, Jpn J. Appl. Phys. \textbf{43},
2779 (2004).

\bibitem{coG2} H. K. Yadalam and U. Harbola, Phys. Rev. B \textbf{94}, 115424 (2016).

\bibitem{maiti} S. K. Maiti, M. Dey, S. Sil, A. Chakrabarti, and S. N. Karmakar,
Europhys. Lett. \textbf{95}, 57008 (2011).

\bibitem{lorentz1} S.-Q. Shen, Phys. Rev. Lett. \textbf{95}, 187203 (2005).

\bibitem{lorentz2} C. J. Kennedy, G. A. Siviloglou, H. Miyake, W. C. Burton, and W. Ketterle,
Phys. Rev. Lett. \textbf{111}, 225301 (2013).

\bibitem{mp16} M. Patra, J. Phys.: Condens. Matter \textbf{34}, 325301 (2022).

\bibitem{fibocharge} M. Patra and S.K. Maiti, Eur. Phys. J. B \textbf{89}, 88 (2016).

\end{thebibliography}
\end{document}